\documentclass[10pt, final, journal, letterpaper, oneside, twocolumn]{IEEEtran}
\usepackage{graphicx}
\usepackage{amsmath}
\usepackage{amssymb}
\usepackage{mathrsfs}
\usepackage{stfloats}
\usepackage{dsfont}
\usepackage{setspace}
\usepackage{array}
\usepackage{bbm}
\usepackage{caption}
\usepackage{epstopdf}
\usepackage[normalem]{ulem}
\usepackage{xcolor}
\usepackage{accents}

\allowdisplaybreaks[4]

\newtheorem{theorem}{\bf{Theorem}}

\newtheorem{proposition}{\bf{Proposition}}
\newtheorem{lemma}{\bf{Lemma}}
\newtheorem{remark}{\bf{Remark}}

\begin{document}
\title{\Large Beamforming Through Reconfigurable Intelligent Surfaces in Single-User MIMO Systems: \\ SNR Distribution and Scaling Laws in the Presence of Channel Fading and Phase Noise}
\author{X.~Qian, M.~Di~Renzo, \IEEEmembership{Fellow,~IEEE}, J.~Liu, A.~Kammoun, \IEEEmembership{Member,~IEEE}, and M.-S.~Alouini, \IEEEmembership{Fellow,~IEEE} \vspace{-0.75cm}

\thanks{Manuscript received May 15, 2020. X. Qian, M. Di Renzo, and J. Liu are with the Laboratoire des Signaux et Syst\`emes, CNRS, CentraleSup\'elec, Universit\'e Paris-Saclay, France. (e-mail: marco.direnzo@centralesupelec.fr). A. Kammoun and M.-S. Alouini are with KAUST, Kingdom of Saudi Arabia. This work was supported by the EC through the H2020 5GstepFWD project under grant 722429 and the H2020 ARIADNE project under grant 871464.}
\thanks{Digital Object Identifier XXX/LWC.2020.XXX}

}
%
%
%
%
\maketitle
\begin{abstract}
We consider a fading channel in which a multi-antenna transmitter communicates with a multi-antenna receiver through a reconfigurable intelligent surface (RIS) that is made of $N$ reconfigurable passive scatterers impaired by phase noise. The beamforming vector at the transmitter, the combining vector at the receiver, and the phase shifts of the $N$ scatterers are optimized in order to maximize the signal-to-noise-ratio (SNR) at the receiver. By assuming Rayleigh fading (or line-of-sight propagation) on the transmitter-RIS link and Rayleigh fading on the RIS-receiver link, we prove that the SNR is a random variable that is equivalent in distribution to the product of three (or two) independent random variables whose distributions are approximated by two (or one) gamma random variables and the sum of two scaled non-central chi-square random variables. The proposed analytical framework allows us to quantify the robustness of RIS-aided transmission to fading channels. For example, we prove that the amount of fading experienced on the transmitter-RIS-receiver channel linearly decreases with $N$. This proves that RISs of large size can be effectively employed to make fading less severe and wireless channels more reliable.
\end{abstract}
%
%
%
%
\vspace{-0.35cm}
\section{Introduction} \label{Introduction}
Reconfigurable intelligent surfaces (RISs) are an emerging transmission technology for application to wireless communications \cite{MDR_Eurasip}. RISs can be realized in different ways, which include (i) implementations based on large arrays of inexpensive antennas that are usually spaced half of the wavelength apart; and (ii) metamaterial-based planar or conformal large surfaces whose scattering elements have sizes and inter-distances much smaller than the wavelength \cite{MDR_JSAC}. In this letter, we consider RISs made of scatterers that are passive, are spaced half of the wavelength apart, and are individually configured and optimized for realizing passive beamforming through the environment \cite{MIT}, \cite{ScatterMIMO}. Compared with other transmission technologies, e.g., phased arrays, multi-antenna transmitters, and relays, RISs require the largest number of scattering elements, but each of them needs to be backed by the fewest and least costly components. Also, no power amplifiers are usually needed. For these reasons, RISs constitute an emerging and promising software-defined architecture that can be realized at reduced cost, size, weight, and power (C-SWaP design) \cite{Pivotal}, \cite{MDR_RISs_Relays}. 

Quantifying the performance of optimized RIS-empowered multi-antenna wireless systems is an open research issue. In particular, several researchers have developed algorithms for jointly optimizing the beamforming vector ($\bf q$) at the transmitter, the matrix of phase shifts at the RIS ($\bf \Phi$), and the combining vector ($\bf w$) at the receiver \cite[Sec. V-J]{MDR_JSAC}. In general, however, the optimal triplet $(\bf q, \bf \Phi, \bf w)$ cannot be formulated in closed-form and can only be computed numerically. An exception is constituted by wireless systems in which the transmitter and receiver are equipped with a single antenna. For this reason, currently available analytical frameworks and scaling laws are only applicable to single-antenna transmitters and receivers. Representative contributions include \cite{Rui_TWC}-\cite{WeiXu_WCL}. In \cite{Rui_TWC} and \cite{Rui_TCOM}, in particular, the authors show that the average SNR at the receiver scales with the square of the number of tunable elements ($N$) of the RIS. In \cite{MDR_Access}, the authors study the error probability over Rayleigh fading channels by using the central limit theorem. In \cite{Justin_WCL} and \cite{Lingyang_TVT}, the authors quantify the impact of phase noise for transmission over Rayleigh and Rician fading channels, respectively. In \cite{WeiXu_WCL}, the authors analyze the impact of phase noise and hardware impairments for transmission over line-of-sight (LOS) channels. 

Motivated by these considerations, we consider a fading channel in which a multi-antenna transmitter communicates with a multi-antenna receiver through an RIS whose $N$ scattering elements are impaired by phase noise. We introduce an analytical approach for characterizing the distribution of the SNR and for determining its scaling laws as a function of $N$. Over Rayleigh fading or LOS channels, we prove that the SNR can be formulated, for any phase noise distribution, as the product of gamma and scaled non-central chi-square random variables. With the aid of numerical simulations, in addition, we show that the SNR can be well approximated with a gamma random variable whose parameters are formulated in closed-form. The proposed approach unveils the scaling laws of the mean, the variance, and the amount of fading (AF) of the SNR as a function of $N$. Our analysis confirms that RISs of large size can be effectively employed to make the transmission of information over fading channels more reliable.

\vspace{-0.35cm}
\section{System Model} \label{SystemModel}
We consider a point-to-point wireless system in which a transmitter equipped with $N_T$ antennas and a receiver equipped with $N_R$ antennas communicate through an RIS. The RIS is made of $N$ antenna elements that are spaced half-wavelength apart and that apply independent phase shifts to the incident signal. The phase shift applied by the $n$th element is denoted by $\phi_n$ for $n=1,2,\ldots,N$. For ease of notation, the $N$ phase shifts are collected in the $N \times N$ diagonal matrix $\bf \Phi$. The $n$th phase shift is assumed to be subject to phase noise, e.g., due to the finite resolution of the phase shifts or to phase estimation errors. The phase noise is assumed to be independent among the $N$ phase shifts. We define ${\phi _n} = \phi _n^{\left( {{\rm{opt}}} \right)} + {\delta _n}$, where $\phi _n^{\left( {{\rm{opt}}} \right)}$ is the optimal phase shift in the absence of phase noise and ${\delta _n}$ is the phase noise. The distribution of ${\delta _n}$ is arbitrary but its mean is assumed to be zero. Examples of phase noise distributions are given in Section \ref{Analysis_SNR}. The $N_T \times 1$ unit-norm beamforming vector at the transmitter is denoted by $\bf q$ and the $N_R \times 1$ unit-norm combining vector at the receiver is denoted by $\bf w$. The triplet $(\bf q, \bf \Phi, \bf w)$ is jointly optimized to maximize the receive SNR. As detailed in Section \ref{ProblemStatement}, we assume that the RIS operates in the far-field regime. Hence, $N$ can be large but cannot tend to infinity \cite[Sec. IV-D]{MDR_JSAC}. The main notation is given in Table \ref{Table_Notation}.

\begin{table}[!t] \footnotesize
\centering
\caption{Main notation (RV = random variable)}
\begin{tabular}{l|l} \hline
\hspace{0.25cm} Symbol & \hspace{2.5cm} Definition \\ \hline
$\mathop  = \limits^d$, $\sim$ & Equivalent in distribution, distributed as  \\
$\mathop  = \limits^{N \gg 1}$, $\mathop  \propto \limits^{N \gg 1}$ & Equality and scaling law if $N \gg 1$ \\
$\mathbb{E}$, $\mathbb{V}$, ${{\rm cov}}$ & Expectation, variance, covariance \\ 
$\rm Re$, $\rm Im$ & Real part, imaginary part \\ 
${\left(  \cdot  \right)^H}$, $\bf{1}{\left(  \cdot  \right)}$ & Hermitian operator, indicator function \\ 
$\left|  \cdot  \right|$, $\left\|  \cdot  \right\|$ & Absolute value, norm of a vector \\
${{\bf{0}}_{A \times B}}$ & $A \times B$ matrix with all zero entries \\
${{\bf{1}}_{A \times B}}$ & $A \times B$ matrix with all one entries \\
$I_{\alpha}(\cdot)$, $\Gamma(\cdot)$ & Bessel function of the first kind, Gamma function \\
${\rm{sinc}}\left( x \right)$ & Normalized sampling function ($\sin \left( {\pi x} \right)/\left( {\pi x} \right)$) \\
$\mathcal{O}\left(  \cdot  \right)$ & Big O (asymptotic) notation \\
${\mathcal{N}}\left( {m,{\sigma ^2}} \right)$ & Gaussian RV ($\mathbb {E} = m$, $\mathbb{V} = \sigma^2$) \\
${\mathcal{CN}}\left( {m,{\sigma ^2}} \right)$ & Complex Gaussian RV ($\mathbb {E} = m$, $\mathbb {V} =\sigma^2$) \\
${\mathcal{B}}\left( {a,b} \right)$ & Beta RV ($\mathbb {E} = a/(a+b)$, $\mathbb {V} = \mathbb{E}(1-\mathbb{E})/(a+b+1)$) \\
${\mathcal{U}}\left( {a,b} \right)$ & Uniform RV in $[a,b]$ \\
${\mathcal{G}}\left( {k,\theta } \right)$ & Gamma RV ($\mathbb {E} = k \theta$, $\mathbb {V} = k \theta^2$) \\
${\mathcal{X}}_k^2\left( \lambda  \right)$ & Non-central chi-square RV ($\mathbb {E} = k+\lambda$, $\mathbb {V} = 2(k+2\lambda)$) \\
${\mathcal{VM}}\left( {\mu ,\kappa } \right)$ & Von Mises RV ($\mathbb {E} = \mu$, $\mathbb {V} = 1 - I_1(\kappa)/I_0(\kappa)$) \\ \hline
\end{tabular}
\label{Table_Notation} \vspace{-0.5cm}
\end{table}

\vspace{-0.35cm}
\subsection{Channel Model} \label{ChannelModel}
The $N_R \times N$ channel matrix on the RIS-receiver link is denoted by $\bf G$. Its entries are assumed to be independent and identically distributed (i.i.d.) complex Gaussian random variables with zero mean and unit variance, i.e., ${\bf{G}} \sim \mathcal{CN} \left( {{{\bf{0}}_{{N_R} \times N}},{{\bf{I}}_{{N_R} \times N}}} \right)$, i.e., Rayleigh fading is considered. This assumption is motivated by the mobility of the receiver and, hence, the difficulty of establishing an LOS link. The $N \times N_T$ channel matrix on the transmitter-RIS link is denoted by $\bf H$. Two canonical case studies are considered for this link. 

\subsubsection{Rayleigh fading} ${\bf{H}} \sim \mathcal{CN} \left( {{{\bf{0}}_{{N} \times N_T}},{{\bf{I}}_{{N} \times N_T}}} \right)$, similar to the RIS-receiver link. This setup is representative of scenarios in which the RISs are randomly deployed, e.g., on spatial blockages whose locations are not under the control of the system designer, and, thus, the locations of the RISs cannot be optimized \cite{MDR_RIS-SG}. Thus, LOS propagation cannot be ensured.

\subsubsection{Deterministic LOS} ${\bf{H}} = \sqrt {{N_T}{N_R}} {{\bf{a}}_{\rm{RIS}}}{\bf{a}}_T^H$, where ${{\bf{a}}_{{T}}}$ and ${{\bf{a}}_{{\rm{RIS}}}}$ are the array responses of the transmitter and RIS, respectively. ${{\bf{a}}_{{T}}}$ is an $N_T \times 1$ unit-norm (i.e., ${\left\| {{{\bf{a}}_T}} \right\|^2} = 1$) vector whose generic entry is ${{\bf{a}}_T}\left( t \right) = \left( 1/ \sqrt{N_T} \right)\exp \left( { - j2\pi f\left( t \right)} \right)$, where $f(t)$ depends on the geometry of the transmit array, and ${{\bf{a}}_{{\rm{RIS}}}}$ is an $N \times 1$ unit-norm (i.e., ${\left\| {{{\bf{a}}_{\rm{RIS}}}} \right\|^2} = 1$) vector whose generic entry is ${{\bf{a}}_{\rm {RIS}}}\left( t \right) = \left( 1/ \sqrt{N} \right)\exp \left( { - j2\pi g\left( n \right)} \right)$, where $g(n)$ depends on the geometry of the RIS. Explicit expressions for ${{\bf{a}}_{{T}}}$ and ${{\bf{a}}_{{\rm{RIS}}}}$ can be found in \cite[Eq. (2)]{WeiXu_WCL}. This setup is representative of scenarios in which the locations of the RISs can be appropriately optimized for ensuring an LOS link \cite{ScatterMIMO}.

The case study in which Rayleigh fading is assumed on both links is denoted by the subscript ``$\rm{RR}$''. The setup in which LOS and Rayleigh fading are assumed on the first and second link, respectively, is denoted by the subscript ``$\rm{LR}$''. The analysis of other channel models is left to future research.

\vspace{-0.35cm}
\section{Problem Formulation} \label{ProblemStatement}
In \cite{MDR_Overhead}, the authors have recently proved that, by jointly optimizing the triplet $(\bf q, \bf \Phi, \bf w)$, the SNR in single-user multi-antenna systems can be tightly approximated as follows:
\begin{equation} \label{Eq_1}
\begin{split}
& {\rm{SN}}{{\rm{R}}_{{\rm{RR}}}} \approx {\gamma _0}{N^2}\mathop {\max }\nolimits_{l,k} \left\{ {{\lambda _{l,{\bf{G}}}}{\lambda _{k,{\bf{H}}}}{\left|  \Upsilon _{l,k} \right|^2 }  } \right\}\\
& {\Upsilon _{l,k}} = { {\sum\nolimits_{n = 1}^N {\left| {{{\bf{v}}_{l,{\bf{G}}}}\left( n \right)} \right|\left| {{{\bf{u}}_{k,{\bf{H}}}}\left( n \right)} \right|\exp \left( {j{\delta _n}} \right)} } }
\end{split}
\end{equation}
\begin{equation} \label{Eq_2}
\begin{split}
& {\rm{SN}}{{\rm{R}}_{{\rm{LR}}}} \approx {\gamma _0} N_T N\mathop {\max }\nolimits_{l} \left\{ {{\lambda _{l,{\bf{G}}}}{\left| \Psi _l \right|^2}} \right\}\\
& {\Psi _l} = {{\sum\nolimits_{n = 1}^N {\left| {{{\bf{v}}_{l,{\bf{G}}}}\left( n \right)} \right|\exp \left( {j{\delta _n}} \right)} } }
\end{split}
\end{equation}
\noindent where: (i) $\gamma_0$ is a scaling factor that accounts for, e.g., the transmission distances, the geometric size of the RIS, the transmission bandwidth, the noise figure \cite{Wankai_Measurements}. In this letter, it is considered to be a constant; (ii) ${{\lambda _{l,{\bf{G}}}}}$ is the $l$th non-zero eigenvalue of the matrix ${{\bf{W}}_{\bf{G}}} = \left( {{1 \mathord{\left/ {\vphantom {1 N}} \right. \kern-\nulldelimiterspace} N}} \right){{\bf{G}}^H}{\bf{G}}$ and ${{\lambda _{l,{\bf{H}}}}}$ is the $l$th non-zero eigenvalue of the matrix ${{\bf{W}}_{\bf{H}}} = \left( {{1 \mathord{\left/ {\vphantom {1 N}} \right. \kern-\nulldelimiterspace} N}} \right){{\bf{H}}}{\bf{H}}^H$; (iii) ${{{\bf{v}}_{l,{\bf{G}}}}}$ and ${{{\bf{u}}_{l,{\bf{G}}}}}$ are the $l$th eigenvectors of ${{\bf{W}}_{\bf{G}}}$ and ${\bf{W}}_{{\bf{G}}^H}$, respectively, that correspond to ${{\lambda _{k,{\bf{G}}}}}$; (iv) and ${{{\bf{v}}_{k,{\bf{H}}}}}$ and ${{{\bf{u}}_{k,{\bf{H}}}}}$ are the $k$th eigenvectors of ${{\bf{W}}_{\bf{H}}}$ and ${\bf{W}}_{{\bf{H}}^H}$, respectively, that correspond to ${{\lambda _{k,{\bf{H}}}}}$. As mentioned, the SNRs in \eqref{Eq_1} and \eqref{Eq_2} are applicable in the far-field regime, as defined in \cite{Wankai_Measurements}, \cite{MDR_SPAWC2020}. Thus, $N$ can be large but it needs to be finite \cite[Sec. IV-D]{MDR_JSAC}.

\vspace{-0.35cm}
\subsection{Preliminaries} \label{Preliminaries}
The semi-analytical expressions of the SNR in \eqref{Eq_1} and \eqref{Eq_2} are the departing point for calculating the distribution and the scaling laws of the SNR as a function of $N$ (see Section \ref{Analysis_SNR}). First, we summarize some lemmas to enable such analysis.
\begin{lemma} \label{Lemma_Eigenvalues}
Let $\lambda _{\bf{H}}^ +$ and $\lambda _{\bf{G}}^ +$ be the largest eigenvalues of ${{\bf{W}}_{\bf{H}}}$ and ${{\bf{W}}_{\bf{G}}}$, respectively. $\lambda _{\bf{H}}^ +$ and $\lambda _{\bf{G}}^ +$ are well approximated by Gamma random variables whose mean and variance are:
\begin{equation} \label{Eq_3}
\begin{split}
& {\mathbb{E}}\left\{ {\lambda _{\bf{X}}^ + } \right\} = {\alpha _1}\left( {M,N} \right) - {\alpha _0}{\beta _1}\left( {M,N} \right)\\
& {\mathop{\mathbb {V}}} \left\{ {\lambda _{\bf{X}}^ + } \right\} = {\beta _0}\beta _1^2\left( {M,N} \right)
\end{split}
\end{equation}
\noindent where ${\bf{X}} = \left\{ {{\bf{G}},{\bf{H}}} \right\}$, $M = N_T$ if ${\bf{X}}={\bf{H}}$ and $M = N_R$ if ${\bf{X}}={\bf{G}}$, ${\alpha _0} = 1.7711$, ${\beta _0} = 0.8132$, and:
\begin{equation} \label{Eq_4}
\begin{split}
& {\alpha _1}\left( {M,N} \right) = {\left( {1 + \sqrt {{M \mathord{\left/
 {\vphantom {M N}} \right.
 \kern-\nulldelimiterspace} N}} } \right)^2}\\
& {\beta _1}\left( {M,N} \right) = {N^{{{ - 2} \mathord{\left/
 {\vphantom {{ - 2} 3}} \right.
 \kern-\nulldelimiterspace} 3}}}\left( {1 + \sqrt {{M \mathord{\left/
 {\vphantom {M N}} \right.
 \kern-\nulldelimiterspace} N}} } \right){\left( {1 + \sqrt {{N \mathord{\left/
 {\vphantom {N M}} \right.
 \kern-\nulldelimiterspace} M}} } \right)^{{1 \mathord{\left/
 {\vphantom {1 3}} \right.
 \kern-\nulldelimiterspace} 3}}}
\end{split}
\end{equation}
\end{lemma}
\begin{IEEEproof}
It follows from \cite{Tirkkonen} and \cite{Chiani} by applying results on random matrix theory and by calculating numerically the mean and the variance of the Tracy-Widom distribution.
\end{IEEEproof} 
\begin{remark}
$\lambda _{\bf{H}}^ +$ and $\lambda _{\bf{G}}^ +$ may be approximated by a shifted Gamma random variable \cite{Chiani}. We consider a Gamma random variable due to its simplicity yet satisfactory accuracy.
\end{remark}
\begin{lemma} \label{Lemma_Eigenvectors}
Let ${{{\bf{v}}_{l,{\bf{G}}}}}$ and ${{{\bf{u}}_{k,{\bf{H}}}}}$ be the eigenvectors in \eqref{Eq_1} and \eqref{Eq_2}. For any $N$, they are i.i.d. and uniformly distributed vectors on the $N-1$ sphere, i.e., on the surface of the unit $N$-ball. Thus, their distribution is equivalent to (for any $l, k$):
\begin{equation} \label{Eq_5}
{{\bf{v}}_{l,{\bf{G}}}}\mathop  = \limits^d {{\bf{v}} \mathord{\left/
 {\vphantom {{\bf{v}} {\left\| {\bf{v}} \right\|}}} \right.
 \kern-\nulldelimiterspace} {\left\| {\bf{v}} \right\|}}\quad \quad \quad {{\bf{u}}_{l,{\bf{H}}}}\mathop  = \limits^d {{\bf{u}} \mathord{\left/
 {\vphantom {{\bf{u}} {\left\| {\bf{u}} \right\|}}} \right.
 \kern-\nulldelimiterspace} {\left\| {\bf{u}} \right\|}}
\end{equation}
\noindent where ${\bf{v}}\left( n \right) \sim {\mathcal{CN}}\left( {0,1} \right)$ and ${\bf{u}}\left( n \right) \sim {\mathcal{CN}}\left( {0,1} \right)$ are mutually i.i.d. random variables for $n=1,2,\ldots,N$.
\end{lemma}
\begin{IEEEproof}
See \cite{Romain_Book}.
\end{IEEEproof}
\begin{remark}
From Lemma \ref{Lemma_Eigenvectors}, we evince that, for every finite $N$, the eigenvectors of a Wishart matrix with zero mean complex Gaussian entries (i.e., ${{\bf{W}}_{\bf{H}}}$ and ${{\bf{W}}_{\bf{G}}}$) do not point towards any privileged direction. If the entries are not Gaussian, this result does not hold in general \cite{Romain_Book}.
\end{remark}
\begin{lemma} \label{Lemma_Eigenvalues_Eigenvectors}
For any $l$, the eigenvalues ${{\lambda _{l,{\bf{X}}}}}$ and the eigenvectors ${{{\bf{v}}_{l,{\bf{X}}}}}$ or ${{{\bf{u}}_{l,{\bf{X}}}}}$ for ${\bf{X}} = \left\{ {{\bf{G}},{\bf{H}}} \right\}$ are independent.
\end{lemma}
\begin{IEEEproof}
See \cite{Romain_Book}.
\end{IEEEproof}
\begin{lemma} \label{Lemma_DistributionSquareAbsEigenvector}
Let ${\bf{y}}$ be an $N \times 1$ vector whose entries are i.i.d. standard complex Gaussian random variables, i.e., ${\bf{y}}\left( n \right) \sim {\mathcal{CN}}\left( {0,1} \right)$ for $n=1,2,\ldots,N$. Define the normalized vector ${\bf{\widehat y}}\left( n \right) = {{{{{\left| {{\bf{y}}\left( n \right)} \right|}}} \mathord{\left/ {\vphantom {{{{\left| {{\bf{y}}\left( n \right)} \right|}^2}} {\left\| {\bf{y}} \right\|}}} \right. \kern-\nulldelimiterspace} {\left\| {\bf{y}} \right\|}}}$. Then, ${\bf{\widehat y}}^2\left( n \right)\sim {\mathcal{B}}\left( {1,N - 1} \right)$ and:
\begin{equation} \label{Eq_6}
\begin{split}
& {\mathbb{E}}\left\{ {{\bf{\widehat y}}^2\left( n \right)} \right\} = {1 \mathord{\left/
 {\vphantom {1 N}} \right.
 \kern-\nulldelimiterspace} N}\\
& {\mathbb{E}}\left\{ { {{\bf{\widehat y}}\left( n \right)} } \right\} = \left( {{{\sqrt \pi  } \mathord{\left/
 {\vphantom {{\sqrt \pi  } 2}} \right.
 \kern-\nulldelimiterspace} 2}} \right)\left( {{{\Gamma \left( N \right)} \mathord{\left/
 {\vphantom {{\Gamma \left( N \right)} {\Gamma \left( {N + {1 \mathord{\left/
 {\vphantom {1 2}} \right.
 \kern-\nulldelimiterspace} 2}} \right)}}} \right.
 \kern-\nulldelimiterspace} {\Gamma \left( {N + {1 \mathord{\left/
 {\vphantom {1 2}} \right.
 \kern-\nulldelimiterspace} 2}} \right)}}} \right)
\end{split}
\end{equation}
\end{lemma}
\begin{IEEEproof}
Since ${\bf{y}}\left( n \right) \sim {\mathcal{CN}}\left( {0,1} \right)$ for $n=1,2,\ldots,N$, then ${\bf{\widehat y}}^2\left( n \right)= {\mathcal{Y}_1 \mathord{\left/ {\vphantom {\mathcal{Y}_1 {\left( {\mathcal{Y}_1 + \mathcal{Y}_2} \right)}}} \right. \kern-\nulldelimiterspace} {\left( {\mathcal{Y}_1 + \mathcal{Y}_2} \right)}}$ where $\mathcal{Y}_1 = {{{\bf{\widehat y}}^2\left( n \right)} } \sim {\mathcal{G}}\left( {1,1} \right)$ and $\mathcal{Y}_2 = \sum\nolimits_{m \ne n = 1}^N {{{ {{\bf{\widehat y}}^2\left( m \right)} }}}  \sim {\mathcal{G}}\left( {N - 1,1} \right)$ are independent random variables. Thus, ${\mathcal{Y}_1 \mathord{\left/ {\vphantom {\mathcal{Y}_1 {\left( {\mathcal{Y}_1 + \mathcal{Y}_2} \right)}}} \right. \kern-\nulldelimiterspace} {\left( {\mathcal{Y}_1 + \mathcal{Y}_2} \right)}} \sim {\mathcal{B}}\left( {1,N - 1} \right)$, and \eqref{Eq_6} follows from the moments of a Beta random variable.
\end{IEEEproof}
\begin{lemma} \label{Lemma_Average_AnAm}
Let ${\bf{y}}$ be an $N \times 1$ vector of i.i.d. standard complex Gaussian random variables, i.e., ${\bf{y}}\left( n \right) \sim {\mathcal{CN}}\left( {0,1} \right)$ for $n=1,2,\ldots,N$. Define ${\bf{\widehat y}}\left( n \right) = {{\left| {{\bf{y}}\left( n \right)} \right|} \mathord{\left/ {\vphantom {{\left| {{\bf{y}}\left( n \right)} \right|} {\left\| {\bf{y}} \right\|}}} \right. \kern-\nulldelimiterspace} {\left\| {\bf{y}} \right\|}}$. For $m \ne n = 1, 2, \ldots, N$, we have ${\mathbb{E}}\left\{ {{\bf{\widehat y}}\left( n \right){\bf{\widehat y}}\left( m \right)} \right\} = {\pi  \mathord{\left/ {\vphantom {\pi  {\left( {4N} \right)}}} \right. \kern-\nulldelimiterspace} {\left( {4N} \right)}}$.
\end{lemma}
\begin{IEEEproof}
Define the variable $z= {\sum\nolimits_{k = 1}^N {\left|{{\bf{y}}}\left( k \right)\right|}^2 }$. By using the notable integral $\int\nolimits_0^{ + \infty } {{e^{ - zt}}dt}  = {1 \mathord{\left/ {\vphantom {1 z}} \right. \kern-\nulldelimiterspace} z}$ for $z >0$, we obtain:
\begin{align} \label{Eq_Extra1}
{\mathbb{E}}\left\{ {{\bf{\widehat y}}\left( n \right){\bf{\widehat y}}\left( m \right)} \right\} &= {\mathbb{E}}\left\{\left| {\bf{y}}\left( n \right)\right| \left|{\bf{y}}\left( m \right)\right|\left( {{1 \mathord{\left/ {\vphantom {1 z}} \right. \kern-\nulldelimiterspace} z}} \right) \right\} \\ 
& \hspace{-1.75cm}\mathop  = \limits^{\left( a \right)} \int\nolimits_0^{ + \infty } {{\mathbb{E}}\left\{ {{{\mathcal{J}}_n}\left( t \right)} \right\}{\mathbb{E}}\left\{ {{{\mathcal{J}}_m}\left( t \right)} \right\}\prod\limits_{k = 1 \ne n,m}^N{\mathbb{E}}\left\{ {{{\mathcal{J}}_k}\left( t \right)} \right\}dt} \nonumber
\end{align}
\noindent where (a) follows because the entries of ${\bf{y}}$ are independent and we defined ${{\mathcal{J}}_n}\left( t \right) = \left|{\bf{ y}}\left( n \right)\right|\exp \left( { -\left| {{{\bf{ y}}}}\left( n \right) \right|^2 t } \right)$, ${{\mathcal{J}}_m}\left( t \right) =$ $\left|{\bf{ y}}\left( m \right)\right|\exp \left( { -\left| {{{\bf{ y}}}}\left( m \right) \right|^2 t } \right)$, ${{\mathcal{J}}_k}\left( t \right) = \exp \left( { - {\left|{{{\bf{ y}}}}\left( k \right)\right|^2} t} \right)$. The proof follows by computing each expectation since the distribution of $\left|{{{\bf{ y}}}}\left( n \right)\right|^2$ is known, i.e., $\left|{{{\bf{ y}}}}\left( n \right)\right|^2 \sim \mathcal{G}\left( {1,1} \right)$, and by using the notable integral $\int\nolimits_0^{ + \infty } {{{\left( {1 + t} \right)}^{1 + N}}dt}  = 1/N$.
\end{IEEEproof}
\begin{lemma} \label{Lemma_Moments_EigevectorSum}
Consider ${{\Upsilon _{l,k}}}$ and ${{\Psi _l}}$ in \eqref{Eq_1} and \eqref{Eq_2} for $l,k=1,2,\ldots,N$. For $\eta=1,2$, let us define the moments $\overline {\overline m}_{\mathop{\mathcal R}\nolimits} ^{\left( \eta  \right)} = {\mathbb{E}}\left\{ {{{\left( {{\mathop{\rm Re}\nolimits} \left\{ {{\Upsilon _{l,k}}} \right\}} \right)}^\eta }} \right\}$, $\overline {\overline m}_{\mathcal{I}}^{\left( \eta  \right)} = {\mathbb{E}}\left\{ {{{\left( {{\mathop{\rm Im}\nolimits} \left\{ {{\Upsilon _{l,k}}} \right\}} \right)}^\eta }} \right\}$, $\overline m_{\mathop{\mathcal R}\nolimits} ^{\left( \eta  \right)} = {\mathbb{E}}\left\{ {{{\left( {{\mathop{\rm Re}\nolimits} \left\{ {{\Psi _l}} \right\}} \right)}^\eta }} \right\}$, and $\overline m_{\mathcal{I}}^{\left( \eta  \right)} = {\mathbb{E}}\left\{ {{{\left( {{\mathop{\rm Im}\nolimits} \left\{ {{\Psi _l}} \right\}} \right)}^\eta }} \right\}$. Then, we have:
\begin{equation} \label{Eq_7}
\begin{split}
& \overline {\overline m}_{\mathop{\mathcal R}\nolimits} ^{\left( 1 \right)} = N\left( {{\pi  \mathord{\left/
 {\vphantom {\pi  4}} \right.
 \kern-\nulldelimiterspace} 4}} \right){\left( {{{\Gamma \left( N \right)} \mathord{\left/
 {\vphantom {{\Gamma \left( N \right)} {\Gamma \left( {N + {1 \mathord{\left/
 {\vphantom {1 2}} \right.
 \kern-\nulldelimiterspace} 2}} \right)}}} \right.
 \kern-\nulldelimiterspace} {\Gamma \left( {N + {1 \mathord{\left/
 {\vphantom {1 2}} \right.
 \kern-\nulldelimiterspace} 2}} \right)}}} \right)^2}{c_1}\\
& \overline {\overline m}_{\mathcal{I}}^{\left( 1 \right)} = N\left( {{\pi  \mathord{\left/
 {\vphantom {\pi  4}} \right.
 \kern-\nulldelimiterspace} 4}} \right){\left( {{{\Gamma \left( N \right)} \mathord{\left/
 {\vphantom {{\Gamma \left( N \right)} {\Gamma \left( {N + {1 \mathord{\left/
 {\vphantom {1 2}} \right.
 \kern-\nulldelimiterspace} 2}} \right)}}} \right.
 \kern-\nulldelimiterspace} {\Gamma \left( {N + {1 \mathord{\left/
 {\vphantom {1 2}} \right.
 \kern-\nulldelimiterspace} 2}} \right)}}} \right)^2}{s_1}\\
& \overline {\overline m}_{\mathop{\mathcal R}\nolimits} ^{\left( 2 \right)} = \left( {{1 \mathord{\left/
 {\vphantom {1 N}} \right.
 \kern-\nulldelimiterspace} N}} \right){c_2} + \left( {{{{\pi ^2}} \mathord{\left/
 {\vphantom {{{\pi ^2}} {16}}} \right.
 \kern-\nulldelimiterspace} {16}}} \right)\left( {{{\left( {N - 1} \right)} \mathord{\left/
 {\vphantom {{\left( {N - 1} \right)} N}} \right.
 \kern-\nulldelimiterspace} N}} \right)c_1^2\\
& \overline {\overline m}_{\mathcal{I}}^{\left( 2 \right)} = \left( {{1 \mathord{\left/
 {\vphantom {1 N}} \right.
 \kern-\nulldelimiterspace} N}} \right){s_2} + \left( {{{{\pi ^2}} \mathord{\left/
 {\vphantom {{{\pi ^2}} {16}}} \right.
 \kern-\nulldelimiterspace} {16}}} \right)\left( {{{\left( {N - 1} \right)} \mathord{\left/
 {\vphantom {{\left( {N - 1} \right)} N}} \right.
 \kern-\nulldelimiterspace} N}} \right)s_1^2
\end{split}
\end{equation}
\begin{equation} \label{Eq_8}
\begin{split}
& \overline m_{\mathcal{R}}^{\left( 1 \right)} = N (\sqrt{\pi/4}) \left( {{{\Gamma \left( N \right)} \mathord{\left/
 {\vphantom {{\Gamma \left( N \right)} {\Gamma \left( {N + {1 \mathord{\left/
 {\vphantom {1 2}} \right.
 \kern-\nulldelimiterspace} 2}} \right)}}} \right.
 \kern-\nulldelimiterspace} {\Gamma \left( {N + {1 \mathord{\left/
 {\vphantom {1 2}} \right.
 \kern-\nulldelimiterspace} 2}} \right)}}} \right){c_1}\\
& \overline m_{\mathcal{I}}^{\left( 1 \right)} = N (\sqrt{\pi/4}) \left( {{{\Gamma \left( N \right)} \mathord{\left/
 {\vphantom {{\Gamma \left( N \right)} {\Gamma \left( {N + {1 \mathord{\left/
 {\vphantom {1 2}} \right.
 \kern-\nulldelimiterspace} 2}} \right)}}} \right.
 \kern-\nulldelimiterspace} {\Gamma \left( {N + {1 \mathord{\left/
 {\vphantom {1 2}} \right.
 \kern-\nulldelimiterspace} 2}} \right)}}} \right){s_1}\\
& \overline m_{\mathcal{R}}^{\left( 2 \right)} = {c_2} + \left( {{\pi  \mathord{\left/
 {\vphantom {\pi  4}} \right.
 \kern-\nulldelimiterspace} 4}} \right)\left( {N - 1} \right)c_1^2\\
& \overline m_{\mathcal{I}}^{\left( 2 \right)} = {s_2} + \left( {{\pi  \mathord{\left/
 {\vphantom {\pi  4}} \right.
 \kern-\nulldelimiterspace} 4}} \right)\left( {N - 1} \right)s_1^2
\end{split}
\end{equation}
\noindent where ${c_1} = {\mathbb{E}}\left\{ {\cos \left( {{\delta _n}} \right)} \right\}$, ${s_1} = {\mathbb{E}}\left\{ {\sin \left( {{\delta _n}} \right)} \right\}$, ${c_2} = {\mathbb{E}}\left\{ {{{\cos }^2}\left( {{\delta _n}} \right)} \right\}$, are ${s_2} = {\mathbb{E}}\left\{ {{{\sin }^2}\left( {{\delta _n}} \right)} \right\}$ are given in Table \ref{Table_II}.
\end{lemma}
\begin{table}[!t] \footnotesize
\centering
\caption{Examples of phase noise distributions ($s_1=0$)}
\newcommand{\tabincell}[2]{\begin{tabular}{@{}#1@{}}#2\end{tabular}}
\begin{tabular}{c|c|c|c} \hline
Distribution & ${c_1}$ & ${c_2}$ & ${s_2}$ \\ \hline
$\delta_n = 0$ & $1$ & $1$ & $0$ \\ 
${\delta _n} \sim {\mathcal{U}}\left( { - \pi ,\pi } \right)$ & $0$ & $1/2$ & $1/2$ \\ 
${\delta _n} \sim {\mathcal{U}}\left( { - \varepsilon \pi ,\varepsilon \pi } \right)$ & ${\rm{sinc}}\left( \varepsilon  \right)$ & $(1+{\rm{sinc}}\left( 2 \varepsilon  \right))/2$ & $(1 - {\rm{sinc}}\left( 2 \varepsilon  \right))/2$ \\ 
${\delta _n} \sim {\mathcal{VM}}\left( { 0 , \kappa } \right)$ & $\frac{{{I_1}\left( \kappa  \right)}}{{{I_0}\left( \kappa  \right)}}$ & $\frac{{{I_1}\left( \kappa  \right) - \kappa {I_2}\left( \kappa  \right)}}{{\kappa {I_0}\left( \kappa  \right)}}$ & $\frac{{{I_1}\left( \kappa  \right)}}{{\kappa {I_0}\left( \kappa  \right)}}$  \\ \hline
\end{tabular}
\label{Table_II} \vspace{-0.5cm}
\end{table}
\begin{IEEEproof}
It follows by re-writing \eqref{Eq_1} and \eqref{Eq_2} by using \eqref{Eq_5}, and by computing the moments using Lemmas \ref{Lemma_DistributionSquareAbsEigenvector} and \ref{Lemma_Average_AnAm}.
\end{IEEEproof}
\begin{lemma} \label{Lemma_Covariance}
Consider ${{\Upsilon _{l,k}}}$ and ${{\Psi _l}}$ in \eqref{Eq_1} and \eqref{Eq_2} for $l,k=1,2,\ldots,N$. The two random variables ${\mathop{\rm Re}\nolimits} \left\{ {{\Upsilon _{l,k}}} \right\}$ and ${\mathop{\rm Im}\nolimits} \left\{ {{\Upsilon _{l,k}}} \right\}$ and the two random variables ${\mathop{\rm Re}\nolimits} \left\{ {{\Psi _l}} \right\}$ and ${\mathop{\rm Im}\nolimits} \left\{ {{\Psi _l}} \right\}$ are uncorrelated for any $l,k=1,2,\ldots,N$ if the distribution of $\delta_n$, for every $n=1,2,\ldots,N$, is symmetric around zero.
\end{lemma}
\begin{IEEEproof}
By definition of covariance, we have:
\begin{equation} \label{Eq_9}
\begin{split}
& {\mathop{\rm cov}} \left\{ {{\mathop{\rm Re}\nolimits} \left\{ {{\Upsilon _{l,k}}} \right\}{\mathop{\rm Im}\nolimits} \left\{ {{\Upsilon _{l,k}}} \right\}} \right\} = \left( {{1 \mathord{\left/
 {\vphantom {1 N}} \right.
 \kern-\nulldelimiterspace} N}} \right){{{\mathbb{E}}\left\{ {\sin \left( {2{\delta _n}} \right)} \right\}} \mathord{\left/
 {\vphantom {{{\mathbb{E}}\left\{ {\sin \left( {2{\delta _n}} \right)} \right\}} 2}} \right.
 \kern-\nulldelimiterspace} 2}\\
& \hspace{1.5cm}+ \left( {{{{\pi ^2}} \mathord{\left/
 {\vphantom {{{\pi ^2}} {16}}} \right.
 \kern-\nulldelimiterspace} {16}}} \right)\left( {{{\left( {N - 1} \right)} \mathord{\left/
 {\vphantom {{\left( {N - 1} \right)} N}} \right.
 \kern-\nulldelimiterspace} N}} \right){c_1}{s_1} - \overline {\overline m}_{\mathop{\mathcal R}\nolimits} ^{\left( 1 \right)}\overline {\overline m}_{\mathcal{I}}^{\left( 1 \right)}
\end{split}
\end{equation}
\begin{equation} \label{Eq_10}
\begin{split}
& {\mathop{\rm cov}} \left\{ {{\mathop{\rm Re}\nolimits} \left\{ {{\Psi _l}} \right\}{\mathop{\rm Im}\nolimits} \left\{ {{\Psi _l}} \right\}} \right\} = {{{\mathbb{E}}\left\{ {\sin \left( {2{\delta _n}} \right)} \right\}} \mathord{\left/
 {\vphantom {{{\mathbb{E}}\left\{ {\sin \left( {2{\delta _n}} \right)} \right\}} 2}} \right.
 \kern-\nulldelimiterspace} 2}\\
& \hspace{1.5cm} + \left( {{\pi  \mathord{\left/
 {\vphantom {\pi  4}} \right.
 \kern-\nulldelimiterspace} 4}} \right)\left( {N - 1} \right){c_1}{s_1} - \overline m_{\mathop{\mathcal R}\nolimits} ^{\left( 1 \right)}\overline m_{\mathcal{I}}^{\left( 1 \right)}
\end{split}
\end{equation}

\noindent The proof follows by noting that ${{\mathbb{E}}\left\{ {\sin \left( {2{\delta _n}} \right)} \right\}}=0$ and $s_1 = 0$ if the distribution of $\delta_n$ is symmetric around zero.
\end{IEEEproof}
\begin{remark}
Based on Table \ref{Table_II}, the distribution of $\delta_n$ is usually symmetric around zero, and the real and imaginary parts of ${{\Upsilon _{l,k}}}$ and ${{\Psi _l}}$ can be assumed to be uncorrelated.
\end{remark}
\begin{remark}
As $N$ grows large, we obtain ${\mathop{\rm cov}} \left\{ {{\mathop{\rm Re}\nolimits} \left\{ {{\Upsilon _{l,k}}} \right\}{\mathop{\rm Im}\nolimits} \left\{ {{\Upsilon _{l,k}}} \right\}} \right\} = \mathcal{O}\left( {{1 \mathord{\left/ {\vphantom {1 N}} \right. \kern-\nulldelimiterspace} N}} \right)$, since, for $N \gg 1$, ${{\Gamma \left( N \right)} \mathord{\left/ {\vphantom {{\Gamma \left( N \right)} {\Gamma \left( {N + {1 \mathord{\left/ {\vphantom {1 2}} \right. \kern-\nulldelimiterspace} 2}} \right)}}} \right. \kern-\nulldelimiterspace} {\Gamma \left( {N + {1 \mathord{\left/ {\vphantom {1 2}} \right. \kern-\nulldelimiterspace} 2}} \right)}} = {N^{ - {1 \mathord{\left/ {\vphantom {1 2}} \right. \kern-\nulldelimiterspace} 2}}}( 1 + {{\left( {8N} \right)}^{ - 1}} + \mathcal{O}\left( {{N^{ - 2}}} \right) )$. This implies that the real and imaginary parts of ${{\Upsilon _{l,k}}}$ are asymptotically (i.e., for large values of $N$) uncorrelated even if the distribution of $\delta_n$ is not symmetric around zero.
\end{remark}
\begin{lemma} \label{Lemma_Distribution_Abs2_EigevectorSum}
Assume that $N$ grows large (i.e., $N \gg 1$). The random variables $\left|{\Upsilon _{l,k}}\right|^2$ and $\left|{\Psi _l}\right|^2$ for $l,k=1,2,\ldots,N$ are (asymptotically) equivalent in distribution to the sum of two scaled non-central chi-square random variables:
\begin{align} \label{Eq_11}
&\left|{\Upsilon _{l,k}}\right|^2 = {\left( {{\mathop{\rm Re}\nolimits} \left\{ {{\Upsilon _{l,k}}} \right\}} \right)^2} + {\left( {{\mathop{\rm Im}\nolimits} \left\{ {{\Upsilon _{l,k}}} \right\}} \right)^2} \mathop  = \limits^{N \gg 1} \overline {\overline \sigma} _{\mathcal R} ^2{{\overline {\overline C}}_{\mathcal{R}}} + \overline {\overline \sigma} _{\mathcal{I}}^2{{\overline {\overline C}}_{\mathcal{I}}} \nonumber \\
& \hspace{1.5cm} {{\overline {\overline C}}_{\mathcal{R}}} \sim \mathcal{X}_1^2\left( {\overline {\overline \mu} _{{\mathcal R}} ^2} \right) \quad {{\overline {\overline C}}_{\mathcal{I}}} \sim \mathcal{X}_1^2\left( {\overline {\overline \mu} _{{\mathcal I}} ^2} \right) 
\end{align} \vspace{-0.5cm}
\begin{align} \label{Eq_12}
& \left|{\Psi _l}\right|^2 =  {\left( {{\mathop{\rm Re}\nolimits} \left\{ {{\Psi _{l}}} \right\}} \right)^2} + {\left( {{\mathop{\rm Im}\nolimits} \left\{ {{\Psi _{l}}} \right\}} \right)^2} \mathop  = \limits^{N \gg 1} \bar \sigma _{\mathop{\mathcal R}\nolimits} ^2{{\overline C}_{\mathcal{R}}} + \overline \sigma _{\mathcal{I}}^2{{\overline C}_{\rm{I}}} \nonumber \\
& \hspace{1.15cm} {{\overline { C}}_{\mathcal{R}}} \sim \mathcal{X}_1^2\left( {\overline { \mu} _{{\mathcal R}} ^2} \right) \quad {{\overline { C}}_{\mathcal{I}}} \sim \mathcal{X}_1^2\left( {\overline { \mu} _{{\mathcal I}} ^2} \right)
\end{align}
\noindent where, for ${\mathcal{S}} = \left\{ {{\mathcal{R}},{\mathcal{I}}} \right\}$, $\overline {\overline \sigma} _{\mathop{\mathcal S}\nolimits} ^2 = \overline {\overline m}_{\mathop{\mathcal S}\nolimits} ^{\left( 2 \right)} - \left( {\overline {\overline m}_{\mathop{\mathcal S}\nolimits} ^{\left( 1 \right)}\overline {\overline m}_{\mathop{\mathcal S}\nolimits} ^{\left( 1 \right)}} \right)$, ${{\overline {\overline \mu }}_{\mathcal{S}}} = {{\overline {\overline m}_{\mathop{\mathcal S}\nolimits} ^{\left( 1 \right)}} \mathord{\left/ {\vphantom {{\overline {\overline m}_{\mathop{\mathcal S}\nolimits} ^{\left( 1 \right)}} {{{\overline {\overline \sigma} }_{\mathcal{S}}}}}} \right. \kern-\nulldelimiterspace} {{{\overline {\overline \sigma} }_{\mathcal{S}}}}}$, and $\overline { \sigma} _{\mathop{\mathcal S}\nolimits} ^2 = \overline { m}_{\mathop{\mathcal S}\nolimits} ^{\left( 2 \right)} - \left( {\overline { m}_{\mathop{\mathcal S}\nolimits} ^{\left( 1 \right)}\overline { m}_{\mathop{\mathcal S}\nolimits} ^{\left( 1 \right)}} \right)$, ${{\overline { \mu }}_{\mathcal{S}}} = {{\overline { m}_{\mathop{\mathcal S}\nolimits} ^{\left( 1 \right)}} \mathord{\left/ {\vphantom {{\overline { m}_{\mathop{\mathcal S}\nolimits} ^{\left( 1 \right)}} {{{\overline { \sigma} }_{\mathcal{S}}}}}} \right. \kern-\nulldelimiterspace} {{{\overline { \sigma} }_{\mathcal{S}}}}}$.
\end{lemma}
\begin{IEEEproof}
It follows from the central limit theorem if $N \gg 1$: ${\mathop{\rm Re}\nolimits} \left\{ {{\Upsilon _{l,k}}} \right\}  \sim  {\mathcal{N}}\left( {\overline {\overline m}_{\mathop{\mathcal R}\nolimits} ^{\left( 1 \right)}, \overline {\overline \sigma} _{\mathop{\mathcal R}\nolimits} ^2} \right)$, ${\mathop{\rm Im}\nolimits} \left\{ {{\Upsilon _{l,k}}} \right\} \sim  {\mathcal{N}}\left( {\overline {\overline m}_{\mathop{\mathcal I}\nolimits} ^{\left( 1 \right)}, \overline {\overline \sigma} _{\mathop{\mathcal I}\nolimits} ^2} \right)$, ${\mathop{\rm Re}\nolimits} \left\{ {{\Psi _{l}}} \right\} \sim {\mathcal{N}}\left( {\overline { m}_{\mathop{\mathcal R}\nolimits} ^{\left( 1 \right)}, \overline { \sigma} _{\mathop{\mathcal R}\nolimits} ^2} \right)$, and ${\mathop{\rm Im}\nolimits} \left\{ {{\Psi _{l}}} \right\} \sim {\mathcal{N}}\left( {\overline { m}_{\mathop{\mathcal I}\nolimits} ^{\left( 1 \right)}, \overline { \sigma} _{\mathop{\mathcal I}\nolimits} ^2} \right)$.
\end{IEEEproof}
\begin{remark}
If the distribution of $\delta_n$ is symmetric around zero, the non-central chi-square random variables in \eqref{Eq_11} and \eqref{Eq_12} are independent. This originates from Remark 3, and because the real and imaginary parts of ${\Upsilon _{l,k}}$ and ${\Psi _{l}}$ converge, asymptotically (i.e., $N \gg 1$), to Gaussian random variables.
\end{remark}
\begin{remark}
Readers are referred to \cite{NonCentral_ChiSquare} for the sum of independent scaled non-central chi-square random variables.
\end{remark}

\vspace{-0.35cm}
\section{Analysis of the Signal-to-Noise-Ratio} \label{Analysis_SNR}
In this section, we analyze the distribution, the mean, the variance, and the AF of the SNR, as well as the corresponding scaling laws as a function of $N$. The AF of the SNR is, in particular, a unified statistical measure that quantifies the severity of fading and, correspondingly, the robustness of transmission technologies against channel fading. Some results are applicable to arbitrary values of $N$, while others apply only for large values of $N$. This is elaborated in further text.

\vspace{-0.35cm}
\subsection{Equivalent in Distribution Representation} \label{EiD}
For an arbitrary $N$, the following theorem yields an equivalent in distribution representation of the SNRs in \eqref{Eq_1} and \eqref{Eq_2}.
\begin{theorem} \label{Theorem_DistributionSNR}
Consider the SNRs in \eqref{Eq_1} and \eqref{Eq_2}. The following equivalent in distribution representations hold true:
\begin{equation} \label{Eq_13}
{\rm{SN}}{{\rm{R}}_{{\rm{RR}}}}\mathop  = \limits^d {\gamma _0}{N^2}\lambda _{\bf{G}}^ + \lambda _{\bf{H}}^ + {\left| {\sum\nolimits_{n = 1}^N {{\bf{\widehat v}}\left( n \right){\bf{\widehat u}}\left( n \right)\exp \left( {j{\delta _n}} \right)} } \right|^2}
\end{equation}
\begin{equation} \label{Eq_14}
{\rm{SN}}{{\rm{R}}_{{\rm{LR}}}}\mathop  = \limits^d {\gamma _0}{N_T}N\lambda _{\bf{G}}^ + {\left| {\sum\nolimits_{n = 1}^N {{\bf{\widehat v}}\left( n \right)\exp \left( {j{\delta _n}} \right)} } \right|^2}
\end{equation}
\end{theorem}
\noindent where ${\bf{\widehat v}}\left( n \right) = {{\left|{\bf{v}}\left( n \right)\right|} \mathord{\left/ {\vphantom {{{\bf{v}}\left( n \right)} {\left\| {\bf{v}} \right\|}}} \right. \kern-\nulldelimiterspace} {\left\| {\bf{v}} \right\|}}$, ${\bf{\widehat u}}\left( n \right) = {{\left|{\bf{u}}\left( n \right)\right|} \mathord{\left/ {\vphantom {{{\bf{u}}\left( n \right)} {\left\| {\bf{u}} \right\|}}} \right. \kern-\nulldelimiterspace} {\left\| {\bf{u}} \right\|}}$, and ${\bf{v}}\left( n \right)$ $\sim {\mathcal{CN}}\left( {0,1} \right)$, ${\bf{u}}\left( n \right) \sim {\mathcal{CN}}\left( {0,1} \right)$ are i.i.d. for $n=1,2,\ldots,N$.
\begin{IEEEproof}
From Lemma \ref{Lemma_Eigenvectors} and Remark 2, the eigenvectors of a Wishart matrix with zero mean complex Gaussian entries are equal in distribution, and, thus, the maximization in \eqref{Eq_1} and \eqref{Eq_2} is determined only by the distribution of the (largest) eigenvalues. From Lemma \ref{Lemma_Eigenvalues_Eigenvectors}, the eigenvectors and the eigenvalues of a Wishart matrix with zero mean complex Gaussian entries are independent. This concludes the proof.
\end{IEEEproof}

Theorem \ref{Theorem_DistributionSNR} provides us with a general tool for the analysis of RIS-aided wireless systems. Let us consider, e.g., ${\rm{SN}}{{\rm{R}}_{{\rm{RR}}}}$. The same comments apply to ${\rm{SN}}{{\rm{R}}_{{\rm{LR}}}}$. Equation \eqref{Eq_13} holds true for any $N$ and it brings to our attention that the SNR is equivalent in distribution to the product of three independent random variables. There exist different approaches for computing the distribution of the product of independent random variables, e.g., \cite{Product_RVs}. For example, the distribution of the square absolute value of the sum in \eqref{Eq_13} may be obtained by first computing the Laplace transform of the sum of independent random variables, which is equal to the product of Laplace transforms of the individual random variables. In this letter, we do not purse this line of research, since the resulting analytical expressions are likely not to be sufficiently tractable to gain insights for system design. In the next two sub-sections, on the other hand, we focus our attention on the case study in which $N \gg 1$, which is relevant for RIS-aided wireless systems.

\begin{table}[!t] \footnotesize
\centering
\caption{SNR scaling laws as a function of $N$ ($s_1=0$). ${{\overline {\overline \zeta} }_{v1}} =  - 6 + \beta_0 \left( {N_T^{ - 1/3} + N_R^{ - 1/3}} \right)$ and ${{\overline {\zeta} }_{v1}} =  - 5 + \beta_0 N_R^{ - 1/3}$.}
\begin{tabular}{l|l} \hline
\hspace{1.0cm} ${\rm{SNR}}_{\rm{RR}}$     &     \hspace{1.0cm}  ${\rm{SNR}}_{\rm{LR}}$ \\ \hline
${{ { o}}_{e0 }} = 1$     &     ${{ { o}}_{e0 }} = N_T$ \\
${{ { o}}_{e1 }} = \left( {{\pi ^2}/16} \right)c_1^2$     &     ${{ { o}}_{e1 }} = \left( {{\pi}/4} \right) N_T c_1^2$ \\ \hline
${{ { o}}_{v0 }} = 2\left( {c_2^2 + s_2^2} \right)$     &     ${{ { o}}_{v0 }} = 2 N_T^2 \left( {c_2^2 + s_2^2} \right)$ \\
$\! \begin{aligned} {{ { o}}_{v1 }} & = \left( {{\pi ^2}/4} \right)c_1^2{c_2} \\ &+ \left( {{\pi ^4}/256} \right){\overline {\overline \zeta} }_{v1}c_1^4 \end{aligned}$     &     $\! \begin{aligned} {{ { o}}_{v1 }} & = {{\pi}} N_T^2 c_1^2{c_2} \\ &+ \left( {{\pi ^4}/16} \right) N_T^2 {\overline { \zeta} }_{v1}c_1^4 \end{aligned}$ \\ \hline
\end{tabular}
\label{Table_ScalingLaws} \vspace{-0.5cm}
\end{table}

\vspace{-0.35cm}
\subsection{Channel Model: Rayleigh Fading -- Rayleigh Fading} \label{Rayleigh_Fading}
In this section, we analyze the statistics of ${\rm{SN}}{{\rm{R}}_{{\rm{RR}}}}$ in \eqref{Eq_13} under the assumption that $N$ is large, i.e., $N \gg 1$.
\begin{theorem} \label{Proposition_RayleighFading_DistributionSNR}
Let us assume the same notation and definitions as in Section \ref{ProblemStatement}. If $N \gg 1$, the following holds true:
\begin{equation} \label{Eq_15}
{\rm{SN}}{{\rm{R}}_{{\rm{RR}}}}\mathop  = \limits^{N \gg 1} {\gamma _0}{N^2}{{{P}}_{\bf{G}}}{{{P}}_{\bf{H}}}\left( {\overline {\overline \sigma} _{\mathop{\mathcal R}\nolimits} ^2{{\overline {\overline C}}_{\mathcal{R}}} + \overline {\overline \sigma} _{\rm{I}}^2{{\overline {\overline C}}_{\mathcal{I}}}} \right)
\end{equation}
\noindent where ${{{P}}_{\bf{X}}} \sim {\mathcal{G}}\left( {{{{{\left( {{\mathbb{E}}\left\{ {\lambda _{\bf{X}}^ + } \right\}} \right)}^2}} \mathord{\left/ {\vphantom {{{{\left( {{\mathbb{E}}\left\{ {\lambda _{\bf{X}}^ + } \right\}} \right)}^2}} {{\mathbb{V}}\left\{ {\lambda _{\bf{X}}^ + } \right\}}}} \right. \kern-\nulldelimiterspace} {{\mathbb{V}}\left\{ {\lambda _{\bf{X}}^ + } \right\}}},{{{\mathbb{V}}\left\{ {\lambda _{\bf{X}}^ + } \right\}} \mathord{\left/ {\vphantom {{{\mathbb{V}}\left\{ {\lambda _{\bf{X}}^ + } \right\}} {{\mathbb{E}}\left\{ {\lambda _{\bf{X}}^ + } \right\}}}} \right. \kern-\nulldelimiterspace} {{\mathbb{E}}\left\{ {\lambda _{\bf{X}}^ + } \right\}}}} \right)$ for ${\bf{X}} = \left\{ {{\bf{G}},{\bf{H}}} \right\}$ and ${{\overline {\overline C}}_{\mathcal{S}}} \sim \mathcal{X}_1^2\left( {\overline {\overline \mu} _{\mathop{\mathcal S}\nolimits} ^2} \right)$ for ${\mathcal{S}} = \left\{ {{\mathcal{R}},{\mathcal{I}}} \right\}$ are four mutually independent random variables.
\end{theorem}
\begin{IEEEproof}
It follows from \eqref{Eq_13} by using Lemmas \ref{Lemma_Eigenvalues} and \ref{Lemma_Distribution_Abs2_EigevectorSum}.
\end{IEEEproof}
\begin{proposition} \label{Proposition_RayleighFading_MeanVarianceSNR}
Let us assume $N \gg 1$. The mean and the variance of ${\rm{SN}}{{\rm{R}}_{{\rm{RR}}}}$ in \eqref{Eq_13} can be formulated as follows:
\begin{equation} \label{Eq_16}
\begin{split}
& {\mathbb{E}}\left\{ {{\rm{SN}}{{\rm{R}}_{{\rm{RR}}}}} \right\}\mathop  = \limits^{N \gg 1} {\gamma _0}{N^2}{{\mathcal{M}}_{\bf{G}}}{{\mathcal{M}}_{\bf{H}}}{\mathcal{\overline {\overline M}}}\\
& {\mathbb{V}}\left\{ {{\rm{SN}}{{\rm{R}}_{{\rm{RR}}}}} \right\}\mathop  = \limits^{N \gg 1} \gamma _0^2{N^4}\left( {{{\mathcal{T}}_{\bf{G}}}{{\mathcal{T}}_{\bf{H}}}{\mathcal{\overline {\overline T}}} - {\mathcal{M}}_{\bf{G}}^2{\mathcal{M}}_{\bf{H}}^2{\mathcal{\overline {\overline M}}}^2} \right)
\end{split}
\end{equation}
\noindent where ${{\mathcal{M}}_{\bf{X}}} = {\mathbb{E}}\left\{ {\lambda _{\bf{X}}^ + } \right\}$ and ${{\mathcal{T}}_{\bf{X}}} = {\mathbb{V}}\left\{ {\lambda _{\bf{X}}^ + } \right\} + {\left( {{\mathbb{E}}\left\{ {\lambda _{\bf{X}}^ + } \right\}} \right)^2}$ for ${\bf{X}} = \left\{ {{\bf{G}},{\bf{H}}} \right\}$, and ${\mathcal{\overline {\overline M}}} = \left( {\overline {\overline \sigma} _{\mathop{\mathcal R}\nolimits} ^2\left( {1 + \overline {\overline \mu} _{\mathop{\mathcal R}\nolimits} ^2} \right) + \overline {\overline \sigma} _{\mathcal{I}}^2\left( {1 + \overline {\overline \mu} _{\mathcal{I}}^2} \right)} \right)$, ${\mathcal{\overline {\overline T}}} = \left( {2 \overline {\overline \sigma} _{\mathop{\mathcal R}\nolimits} ^4\left( {1 + 2 \overline {\overline \mu} _{\mathop{\mathcal R}\nolimits} ^2} \right) + 2 \overline {\overline \sigma} _{\mathcal{I}}^4\left( {1 + 2 \overline {\overline \mu} _{\mathcal{I}}^2} \right)} \right)$.
\end{proposition}
\begin{IEEEproof}
It follows from the independence of the random variables in \eqref{Eq_15} and by using Lemma \ref{Lemma_Eigenvalues} and Lemma \ref{Lemma_Distribution_Abs2_EigevectorSum}.
\end{IEEEproof}

\vspace{-0.35cm}
\subsection{Channel Model: Line-of-Sight -- Rayleigh Fading} \label{LOS_Fading}
In this section, we analyze the statistics of ${\rm{SN}}{{\rm{R}}_{{\rm{LR}}}}$ in \eqref{Eq_14} under the assumption that $N$ is large, i.e., $N \gg 1$.
\begin{theorem} \label{Proposition_LOS_DistributionSNR}
Let us assume the same notation and definitions as in Section \ref{ProblemStatement}. If $N \gg 1$, the following holds true:
\begin{equation} \label{Eq_18}
{\rm{SN}}{{\rm{R}}_{{\rm{LR}}}}\mathop  = \limits^{N \gg 1} {\gamma _0}{N_T N}{{{P}}_{\bf{G}}}\left( {\overline { \sigma} _{\mathop{\mathcal R}\nolimits} ^2{{\overline { C}}_{\mathcal{R}}} + \overline { \sigma} _{\rm{I}}^2{{\overline { C}}_{\mathcal{I}}}} \right)
\end{equation}
\noindent where ${{{P}}_{\bf{G}}} \sim {\mathcal{G}}\left( {{{{{\left( {{\mathbb{E}}\left\{ {\lambda _{\bf{G}}^ + } \right\}} \right)}^2}} \mathord{\left/ {\vphantom {{{{\left( {{\mathbb{E}}\left\{ {\lambda _{\bf{G}}^ + } \right\}} \right)}^2}} {{\mathbb{V}}\left\{ {\lambda _{\bf{G}}^ + } \right\}}}} \right. \kern-\nulldelimiterspace} {{\mathbb{V}}\left\{ {\lambda _{\bf{G}}^ + } \right\}}},{{{\mathbb{V}}\left\{ {\lambda _{\bf{G}}^ + } \right\}} \mathord{\left/ {\vphantom {{{\mathbb{V}}\left\{ {\lambda _{\bf{G}}^ + } \right\}} {{\mathbb{E}}\left\{ {\lambda _{\bf{G}}^ + } \right\}}}} \right. \kern-\nulldelimiterspace} {{\mathbb{E}}\left\{ {\lambda _{\bf{G}}^ + } \right\}}}} \right)$ and ${{\overline { C}}_{\mathcal{S}}} \sim \mathcal{X}_1^2\left( {\overline {\overline \mu} _{\mathop{\mathcal S}\nolimits} ^2} \right)$ for ${\mathcal{S}} = \left\{ {{\mathcal{R}},{\mathcal{I}}} \right\}$ are three mutually independent random variables.
\end{theorem}
\begin{IEEEproof}
It is similar to the proof of Theorem \ref{Proposition_RayleighFading_DistributionSNR}.
\end{IEEEproof}
\begin{proposition} \label{Proposition_LOS_MeanVarianceSNR}
Let us assume $N \gg 1$. The mean and the variance of ${\rm{SN}}{{\rm{R}}_{{\rm{LR}}}}$ in \eqref{Eq_14} can be formulated as follows:
\begin{equation} \label{Eq_19}
\begin{split}
& {\mathbb{E}}\left\{ {{\rm{SN}}{{\rm{R}}_{{\rm{LR}}}}} \right\}\mathop  = \limits^{N \gg 1} {\gamma _0}{N_T N}{{\mathcal{M}}_{\bf{G}}}{\mathcal{\overline { M}}}\\
& {\mathbb{V}}\left\{ {{\rm{SN}}{{\rm{R}}_{{\rm{LR}}}}} \right\}\mathop  = \limits^{N \gg 1} \gamma _0^2{N_T^2 N^2}\left( {{{\mathcal{T}}_{\bf{G}}}{\mathcal{\overline { T}}} - {\mathcal{M}}_{\bf{G}}^2{\mathcal{\overline { M}}}^2} \right)
\end{split}
\end{equation}
\noindent where ${{\mathcal{M}}_{\bf{G}}} = {\mathbb{E}}\left\{ {\lambda _{\bf{G}}^ + } \right\}$, ${{\mathcal{T}}_{\bf{G}}} = {\mathbb{V}}\left\{ {\lambda _{\bf{G}}^ + } \right\} + {\left( {{\mathbb{E}}\left\{ {\lambda _{\bf{G}}^ + } \right\}} \right)^2}$, ${\mathcal{\overline { M}}} = \left( {\overline { \sigma} _{\mathop{\mathcal R}\nolimits} ^2\left( {1 + \overline { \mu} _{\mathop{\mathcal R}\nolimits} ^2} \right) + \overline { \sigma} _{\mathcal{I}}^2\left( {1 + \overline { \mu} _{\mathcal{I}}^2} \right)} \right)$, and ${\mathcal{ {\overline T}}} = \left( {2 \overline { \sigma} _{\mathop{\mathcal R}\nolimits} ^4\left( {1 + 2 \overline { \mu} _{\mathop{\mathcal R}\nolimits} ^2} \right) + 2 \overline { \sigma} _{\mathcal{I}}^4\left( {1 + 2 \overline { \mu} _{\mathcal{I}}^2} \right)} \right)$.
\end{proposition}
\begin{IEEEproof}
It is similar to the proof of Proposition \ref{Proposition_RayleighFading_MeanVarianceSNR}.
\end{IEEEproof}

\vspace{-0.35cm}
\subsection{Scaling Laws and Insights} \label{Insights}
From Propositions \ref{Proposition_RayleighFading_MeanVarianceSNR} and \ref{Proposition_LOS_MeanVarianceSNR}, explicit analytical expressions for the mean and the variance of the SNR can be obtained. The resulting formulas are, however, not tractable enough to gain insights for system design. Therefore, we analyze the dominant terms (scaling laws) in the asymptotic regime $N \gg 1$. 

\begin{proposition} \label{Proposition_RayleighFading_ScalingLawsSNR}
Define ${\rm{D}} = \left\{ {{\rm{RR}},{\rm{LR}}} \right\}$,  and assume $s_1=0$ and $N \gg 1$. Let ${\rm{A}}{{\rm{F}}_{{\rm{SN}}{{\rm{R}}_{{\rm{D}}}}}} = {{{\mathbb{V}}\left\{ {{\rm{SN}}{{\rm{R}}_{{\rm{D}}}}} \right\}} \mathord{\left/ {\vphantom {{{\mathbb{V}}\left\{ {{\rm{SN}}{{\rm{R}}_{{\rm{D}}}}} \right\}} {{{\left( {{\mathbb{E}}\left\{ {{\rm{SN}}{{\rm{R}}_{{\rm{D}}}}} \right\}} \right)}^2}}}} \right. \kern-\nulldelimiterspace} {{{\left( {{\mathbb{E}}\left\{ {{\rm{SN}}{{\rm{R}}_{{\rm{D}}}}} \right\}} \right)}^2}}}$ be the AF of ${\rm{SN}}{{\rm{R}}_{{\rm{D}}}}$. As a function of $N$, while keeping the other system parameters fixed, the following scaling laws hold true:
\begin{align} \label{Eq_17}
& {\mathbb{E}}\left\{ {{\rm{SN}}{{\rm{R}}_{{\rm{D}}}}} \right\}\mathop  \propto \limits^{N \gg 1} {{ { o}}_{e0 }}{N^{1}}{\bf{1}}\left( {{c_1} = 0} \right) + {{ { o}}_{e1 }}{N^{2}}{\bf{1}}\left( {{c_1} \ne 0} \right) \nonumber \\
& {\mathbb{V}}\left\{ {{\rm{SN}}{{\rm{R}}_{{\rm{D}}}}} \right\}\mathop  \propto \limits^{N \gg 1} {{ { o}}_{v0}}{N^{2}}{\bf{1}}\left( {{c_1} = 0} \right) + {{ { o}}_{v1}}{N^{3}}{\bf{1}}\left( {{c_1} \ne 0} \right)\\
& {\rm{A}}{{\rm{F}}_{{\rm{SN}}{{\rm{R}}_{{\rm{D}}}}}}\mathop  \propto \limits^{N \gg 1} \frac{{ { o}}_{v0}} {{ { o}}_{e0}^2} {N^{0}}{\bf{1}}\left( {{c_1} = 0} \right) + \frac{{ { o}}_{v1}} {{ { o}}_{e1}^2}{N^{-1}}{\bf{1}}\left( {{c_1} \ne 0} \right) \nonumber
\end{align}
\noindent where ${{ { o}}_{e0}}$, ${{ { o}}_{e1}}$, ${{ { o}}_{v0}}$, and ${{ { o}}_{v1}}$ are defined in Table \ref{Table_ScalingLaws}.
\end{proposition}
\begin{IEEEproof}
It follows from \eqref{Eq_16}, \eqref{Eq_19}, \eqref{Eq_4} since ${\alpha _1}\left( {M,N} \right) = 1 + {\mathcal{O}}\left( {{N^{ - {1 \mathord{\left/ {\vphantom {1 2}} \right. \kern-\nulldelimiterspace} 2}}}} \right)$, ${\beta _1}\left( {M,N} \right) = {M^{ - {1 \mathord{\left/ {\vphantom {1 6}} \right. \kern-\nulldelimiterspace} 6}}}{N^{ - {1 \mathord{\left/ {\vphantom {1 2}} \right. \kern-\nulldelimiterspace} 2}}} + {\mathcal{O}}\left( {{N^{ - 1}}} \right)$.
\end{IEEEproof}

From Proposition \ref{Proposition_RayleighFading_ScalingLawsSNR}, we can draw the following conclusions on the scaling laws of the SNR as a function of $N$.\\
{\bf{--}} The scaling laws highly depend on whether $c_1=0$ or $c_1 \ne 0$. From Table \ref{Table_II}, e.g., the condition $c_1 = 0$ corresponds to the case study of totally random phase noise. Also, the condition $c_1 = 0$ can be thought of as representative of a scenario with no controllable RIS, in which the RIS is a conventional wall whose phase response is unknown and cannot be optimized. If $c_1=0$, in particular, the AF is constant with $N$, since the RIS is not capable of customizing the radio waves. If $c_1 \ne 0$, on the other hand, the AF decays linearly with $N$. This unveils the capability of RISs of reducing the fading severity and, as a result, making the transmission of information more robust. \\
{\bf{--}} The robustness of RISs against the phase noise can be quantified by studying the ratios ${{{o_{e1}}\left( {{c_1}} \right)} \mathord{\left/ {\vphantom {{{o_{e1}}\left( {{c_1}} \right)} {{o_{e1}}\left( {{c_1} = 1} \right)}}} \right. \kern-\nulldelimiterspace} {{o_{e1}}\left( {{c_1} = 1} \right)}}$ and ${{{o_{v1}}\left( {{c_1},{c_2}} \right)} \mathord{\left/
 {\vphantom {{{o_{v1}}\left( {{c_1},{c_2}} \right)} {{o_{v1}}\left( {{c_1} = 1,{c_2} = 1} \right)}}} \right.
 \kern-\nulldelimiterspace} {{o_{v1}}\left( {{c_1} = 1,{c_2} = 1} \right)}}$ defined in Table \ref{Table_ScalingLaws}, since $c_1=c_2=1$ for the benchmark setup with no phase noise (see Table \ref{Table_II}). This provides a simple tool for quantifying, e.g., the discretization of the phase shifts that yields a suitable trade-off between performance and implementation complexity.

\begin{figure}[!t]
\centering
\includegraphics[width=0.90\columnwidth]{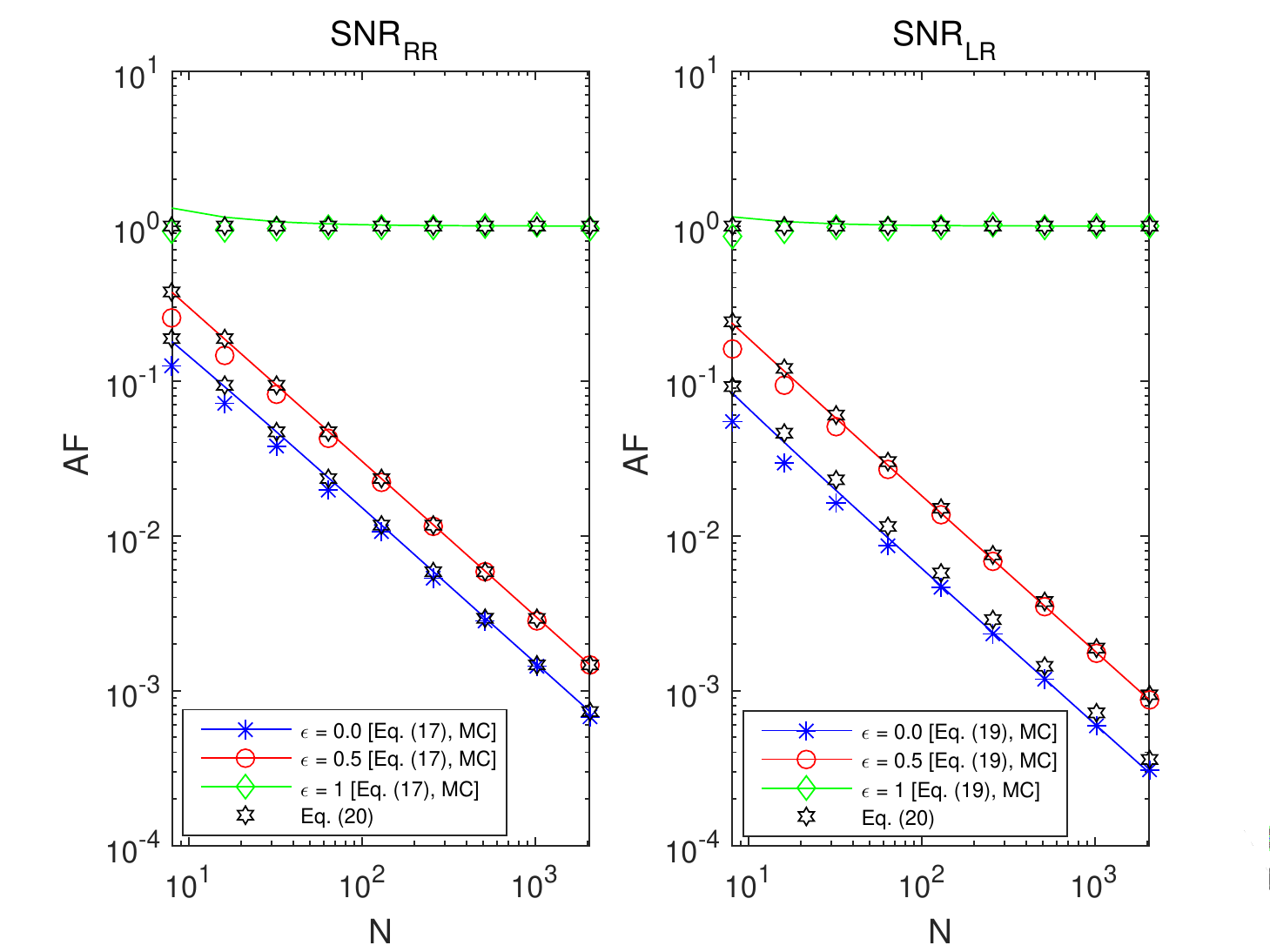}
\vspace{-0.25cm} \caption{AF ($\gamma_0 = 1$, $N_T=N_R=4$, ${\delta _n} \sim {\mathcal{U}}\left( { - \varepsilon \pi ,\varepsilon \pi } \right)$).}
\label{Fig_1} \vspace{-0.5cm}
\end{figure}
\begin{figure}[!t]
\centering
\includegraphics[width=0.90\columnwidth]{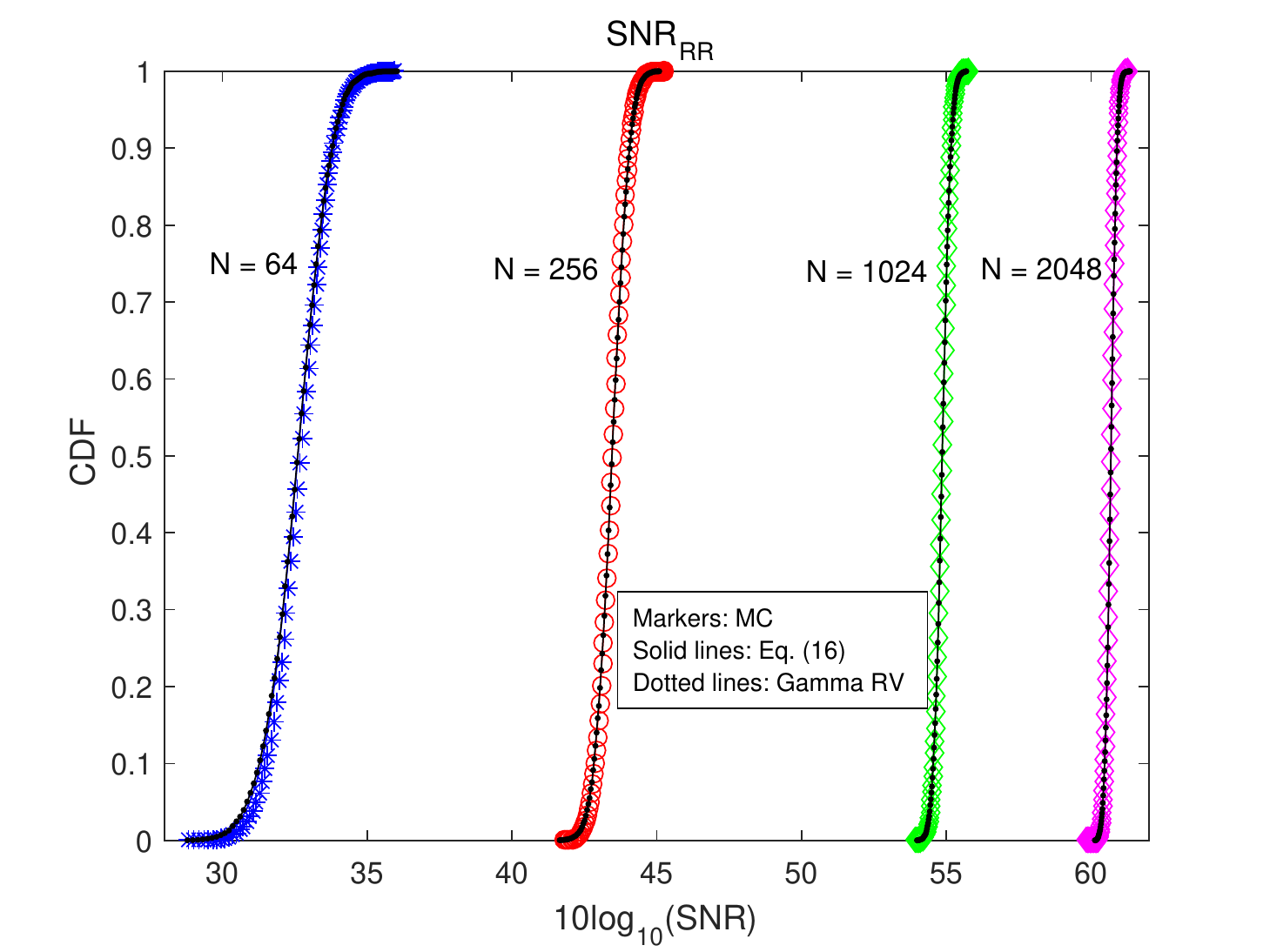}
\vspace{-0.25cm} \caption{CDF ($\gamma_0 = 1$, $N_T=N_R=4$, ${\delta _n} \sim {\mathcal{U}}\left( { - \varepsilon \pi ,\varepsilon \pi } \right)$).}
\label{Fig_2} \vspace{-0.5cm}
\end{figure}

\vspace{-0.35cm}
\section{Numerical Results} \label{NumericalResults} \vspace{-0.15cm}
Figure \ref{Fig_1} shows the AF obtained from Monte Carlo (MC) simulations \cite[Proposition 1]{MDR_Overhead} (markers), and compares it against the analytical frameworks in \eqref{Eq_16} and \eqref{Eq_19} (solid lines), and the scaling laws in \eqref{Eq_17}. Figure \ref{Fig_2} shows the cumulative distribution function (CDF) of the SNR obtained from Monte Carlo simulations \cite[Proposition 1]{MDR_Overhead}, and compares it against the distributions in \eqref{Eq_15}, and a Gamma-based approximation for the SNR, i.e., ${{\rm{P}}_{{\rm{SN}}{{\rm{R}}_{\rm{RR}}}}} \sim {\mathcal{G}}\left( {{{{{\left( {{\mathbb{E}}\left\{ {{\rm{SN}}{{\rm{R}}_{\rm{RR}}}} \right\}} \right)}^2}} \mathord{\left/ {\vphantom {{{{\left( {{\mathbb{E}}\left\{ {{\rm{SN}}{{\rm{R}}_{\rm{RR}}}} \right\}} \right)}^2}} {{\mathbb{V}}\left\{ {{\rm{SN}}{{\rm{R}}_{\rm{RR}}}} \right\}}}} \right. \kern-\nulldelimiterspace} {{\mathbb{V}}\left\{ {{\rm{SN}}{{\rm{R}}_{\rm{RR}}}} \right\}}},{{{\mathbb{V}}\left\{ {{\rm{SN}}{{\rm{R}}_{\rm{RR}}}} \right\}} \mathord{\left/ {\vphantom {{{\rm{V}}\left\{ {{\rm{SN}}{{\rm{R}}_{\rm{D}}}} \right\}} {{\mathbb{E}}\left\{ {{\rm{SN}}{{\rm{R}}_{\rm{RR}}}} \right\}}}} \right. \kern-\nulldelimiterspace} {{\mathbb{E}}\left\{ {{\rm{SN}}{{\rm{R}}_{\rm{RR}}}} \right\}}}} \right)$. The proposed analytical approach is in good agreement with the simulations and confirm our findings.

\vspace{-0.20cm}
\section{Conclusion} \label{Conclusion} \vspace{-0.15cm}
We have introduced an analytical framework to quantify the performance of RIS-aided multi-antenna transmission. If $N \gg 1$, we have proved that the AF of the SNR linearly decreases with $N$. Also, we have shown that the distribution of the SNR can be well approximated with a Gamma random variable. The proposed approach can be generalized to multiple scenarios, e.g., the analysis of multi-user and multi-RIS transmission.

\bibliographystyle{IEEEtran}

\end{document}